\documentclass[journal=jctc,manuscript=article]{achemso}

\usepackage{chemformula} % Formula subscripts using \ch{}
\usepackage[T1]{fontenc} % Use modern font encodings
\usepackage{hyperref}       % hyperlinks
\usepackage{graphicx}       % pictures
\usepackage{url}            % simple URL typesetting
\usepackage{booktabs}       % professional-quality tables
\usepackage{amsfonts}       % blackboard math symbols
\usepackage{nicefrac}       % compact symbols for 1/2, etc.
\usepackage{microtype}      % microtypography
\usepackage{comment}
\usepackage{natbib}
\usepackage{subcaption}
\usepackage{standalone}
\usepackage{amsmath}
\usepackage{algorithm2e}
\usepackage{import}
\usepackage{cleveref}
\usepackage{listings}
\usepackage{color}
\definecolor{deepblue}{rgb}{0,0,0.5}
\definecolor{deepred}{rgb}{0.6,0,0}
\definecolor{deepgreen}{rgb}{0,0.5,0}

\lstdefinestyle{mystyle}{
 keywords={typeof, null, catch, switch, in, int, str, float, self},
 keywordstyle=\color{deepgreen}\bfseries,
 ndkeywords={boolean, throw, import},
 ndkeywords={return, class, if ,elif, endif, while, do, else, True, False , catch, def},
 ndkeywordstyle=\color{black}\bfseries,
 identifierstyle=\color{black},
 sensitive=false,
 comment=[l]{\#},
 morecomment=[s]{/*}{*/},
 commentstyle=\color{deepblue}\ttfamily,
 stringstyle=\color{deepred}\ttfamily,
 breaklines=true,
 captionpos=b,
}

\newcommand{\UCB}{AdaptiveBandit}

\DeclareMathOperator*{\argmax}{argmax}
\author{Adri\`a P\'erez }

\author{Pablo Herrera-Nieto}
\affiliation[UPF]
{Computational Science Laboratory, Universitat Pompeu Fabra, Barcelona, Spain}

\author{Stefan Doerr}
\affiliation[Acellera]
{Acellera Ltd., Barcelona, Spain}
\author{Gianni De Fabritiis}
\email{gianni.defabritiis@upf.edu}
\affiliation[UPF]
{Computational Science Laboratory, Universitat Pompeu Fabra, Barcelona, Spain}
\alsoaffiliation[Acellera]
{Acellera Ltd., Barcelona, Spain}
\alsoaffiliation[ICREA]{Instituci\'o Catalana de Recerca i Estudis Avan\c{c}ats, Barcelona, Spain}

\title{AdaptiveBandit: A multi-armed bandit framework for adaptive sampling in molecular simulations}

\abbreviations{IR,NMR,UV}

\begin{document}


\begin{abstract}
Sampling from the equilibrium distribution has always been a major problem in molecular simulations due to the very high dimensionality of conformational space.  Over several decades, many approaches have been used to overcome the problem. In particular, we focus on unbiased simulation methods such as parallel and adaptive sampling. Here, we recast  adaptive sampling schemes on the basis of multi-armed bandits and develop a novel adaptive sampling algorithm under this framework, \UCB. We test it on multiple simplified potentials and in a protein folding scenario. We find that this framework performs similarly or better in every type of test potentials compared to previous methods. Furthermore, it provides a novel framework to develop new sampling algorithms with better asymptotic characteristics. 
\end{abstract}

\section{Introduction}
In computational biology, macroscopic measurements  by computer simulations are obtained by simulating microscopic molecular systems made of the order of a hundred thousand degrees of freedom. Statistical mechanics tells us  what is the analytical form of the equilibrium distribution given the macroscopic constraint of the environment, e.g. constant temperature, pressure, and number of atoms. Therefore the problem consists in generating samples from such distribution. 

Molecular simulation methods have always been hampered by sampling limitations over the equilibrium distribution due to their computational cost\cite{martinez2017drug, perez2018simulations}. The two main forms to obtain samples are molecular dynamics (MD), a numerical scheme where the propagator of the dynamical system is discretized in time and iterated for billions of steps, and Monte Carlo sampling (MC), where the Monte Carlo rule is used to draw samples from the distribution. These sampling methods are also commonly used in  other fields to sample for arbitrary probability distributions, and many of the methods developed for molecular simulations have been exploited in such contexts later, for instance, umbrella sampling \cite{torrie1977nonphysical}, biased Montecarlo methods \cite{Frenkel:1996:UMS:547952} or biased molecular dynamics like replica-exchange\cite{sugita1999replica, fukunishi2002hamiltonian}, steered MD \cite{izrailev1999steered,isralewitz2001steered}, metadynamics \cite{laio2002escaping}, etc. Progress in molecular simulation sampling has therefore shown its relevance to a broader field of problems. Recently, a new generative method based on normalizing flows \cite{rezende2015variational} has been proposed to sample from the Boltzmann distribution \cite{noe2019boltzmann}.

Due to the difficulties in determining the bias a priori, practically equivalent to having a good prior, unbiased methods such as adaptive sampling \cite{singhal2005error, hinrichs2007calculation,pronk2011copernicus,doerr2014fly} have been recently developed and used successfully \cite{noe2009constructing, plattner2017complete}. Equally, due to the difficulty in generating good Montecarlo moves, molecular dynamics is  almost always preferred to Montecarlo methods, largely due to the current efficiency of generating trajectories rooted in the capability of modern hardware. Specialized computer chips like Anton \cite{shaw2008anton} made possible to run long simulations of the order of hundreds of microseconds, sampling reversibly fast processes and exploring longer timescales \cite{lindorff2011fast}. The advent of GPUs and GPU molecular dynamics software \cite{friedrichs2009accelerating, harvey2009implementation,harvey2009acemd, Eastman2017} was a notable improvement, greatly increasing the computational efficiency of simulations. This, combined with Markov state models (MSMs) \cite{prinz2011markov,bowman2013introduction} allowed to reconstruct a complete statistical description of the full dynamical system from many shorter trajectories, obtaining a description that is equivalent to reversible sampling, once at convergence. 

Running not one, but hundreds or thousands of simulation trajectories \cite{buch2011complete,martinez2018molecular} created a new opportunity to decide the starting conditions of these simulations to obtain the best equilibrium characterization at the minimal computational cost, i.e. adaptive sampling. Initially, adaptive sampling algorithms \cite{singhal2005error, doerr2014fly} were used to reduce statistical uncertainty by choosing conformations that contributed  the most to the error in mean first passage time of an MSM \cite{singhal2005error}, eigenvalues, and eigenvectors \cite{hinrichs2007calculation}, or choosing low state populations \cite{pronk2011copernicus,doerr2014fly}. Furthermore, similar algorithms appeared recently which introduced prior knowledge to the selection criteria \cite{sabbadin2014supervised,perez2015accelerating,zimmerman2015fast}, seeking to further speed up sampling towards equilibrium. One notable example is where contact information is used for protein folding \cite{ovchinnikov2017protein} or bound state contacts in protein-ligand or protein-protein binding \cite{plattner2017complete}. Other applications have used alternative geometric features, such as RMSD or pocket volume, to improve conformational exploration \cite{zimmerman2017prediction} and to find cryptic pockets \cite{cruz2020discovery}. In general, the adaptive sampling policy was always empirical, not based on any mathematical decision process, even though similarities have been recognized with the multi-armed bandit problem \cite{zimmerman2015fast, zimmerman2018choice} and reinforcement learning \cite{shamsi2018reinforcement} before. 

Here we frame adaptive sampling in terms of a multi-armed bandit problem and propose \UCB, an algorithm that uses an action-value function and an upper confidence bound \cite{lai1985asymptotically, auer2002using} selection algorithm, improving adaptive sampling's performance and increasing its versatility when faced against different free energy landscapes. Our main goal is to provide strong fundamentals when facing the exploration-exploitation dilemma by redefining it in terms of reinforcement learning, creating a solid framework from where to easily develop novel algorithms. \UCB\  is available in HTMD (\url{https://github.com/Acellera/htmd}) \cite{doerr2016htmd}.

\section{Methods}
\subsection{MD Simulations}

The configurational space of a molecular system for MD simulations is given by $\chi =\{x=(\mathbf{r}_1,\dots,\mathbf{r}_N) \in \mathbb{R}^{3N}\}$, where N is the number of atoms of the system. Experimental observables $O$ are measured as equilibrium expectations $<O> \,= \int O(x)\,\mu(x)\, dx$, where $\mu(x)$ is the equilibrium distribution. The form of this distribution is known, for instance, the Boltzmann distribution in the canonical ensemble at temperature T is
\begin{equation}\label{boltzmann}
  \mu(x)=e^{\frac{-U(x)}{k_BT}} ,
\end{equation}
where $U(x)$ is the molecular potential energy and $k_BT$ is the Boltzmann constant multiplied by the temperature. MD numerically solves  Newton's equation over the potential $U(x)$ for the variable $x$, plus a Langevin stochastic term accounting for thermal fluctuations \cite{loncharich1992langevin}. Now consider the state $x(t)\in\chi$ as a specific conformation inside the configurational space $\chi$ at time $t$, the probability of finding the molecule in configuration $x_{t+\tau}$ at a later time can be expressed by the conditional transition density function  $p_\tau$, $x_{t+\tau} \sim p_{\tau}(x_{t+\tau} | x_t)$
which describes the probability of finding state $x_{t+\tau}$ given state $x_t$ at time $t$ after a time increment $\tau$.  When performing an MD simulation, the dynamics of the molecular system propagates the state $x_t$ across time. Therefore, MD samples from the transition density $p_\tau$ given discrete time-steps $\tau$ to obtain the next state $x_{t+\tau}$. The process is repeated for many steps, generating a trajectory of conformations. 

The main goal when performing MD simulations is to obtain a good representation of the system's equilibrium distribution $\mu(x)$ i.e. the probability to find conformation $x$ under equilibrium conditions, in order to measure the average of observable $<O>$. If an MD trajectory $\tau$ is long enough, sampling from $p_\tau$ is equivalent to sampling from $\mu(x)$ (\Cref{boltzmann})
\begin{equation}
    \lim_{\tau\to\infty} p_{\tau}(x_{t+\tau} | x_t) = \mu(x) \label{eqsim} .
\end{equation}

Generating long enough trajectories is computationally expensive, and often practically impossible when trying to sample slow events. However, long trajectories can be  substituted by short parallelized trajectories. While in principle one could model directly the conditional probability in \Cref{eqsim}, in practice this is not possible given the very high dimensional space. Fortunately, it can be shown that the dynamics can be separated into  a slow and fast set of variables \cite{prinz2011markov}, and because contributions of fast variables decay exponentially in $\tau$, a reliable MSM can be constructed in terms of the slow variables to compute thermodynamic averages. Usually, time-independent component analysis (tICA) \cite{perez2013identification} and clustering methods are used to learn this set of variables during sampling, necessary to build the MSM.  
Once we obtain the MSM, computed by estimating transition probabilities from discrete conformational states, one can derive thermodynamic and kinetic properties, just assuming local, not global, equilibrium (i.e. $\tau$ is much shorter than what is necessary to satisfy \Cref{eqsim}). 
 
\subsection{The multi-armed bandit problem}

The multi-armed bandit problem is a simplified reinforcement learning setting where one faces the exploration versus exploitation dilemma. 
The problem is defined as a tuple $\langle\mathcal{A},\mathcal{R},\gamma\rangle$, where $\mathcal{A}$ is a set of $k$ actions $\mathcal{A} = \{a_1, a_2,\dots,a_k\}$  and $\mathcal{R}$ is an unknown probability distribution $\mathcal{R}^a = \mathbb{P}[r|a]$ of rewards given the chosen action. We choose $\gamma=0$ for totally discounted rewards. At each time-step $t$, the agent applies a policy $\pi_a=\mathbb{P}[a]$ to select an action $a_t \in \mathcal{A}$, based on previous actions taken and the respectively obtained rewards. Subsequently, the environment returns a reward $r_t \sim R^{a_t}$. Given that we set $\gamma=0$, we define the value of an action $Q_\pi(a)$  as its instantaneous mean reward
\begin{equation}
    Q_\pi(a) = \mathbb{E}_\pi [r | a] .
\end{equation}

The goal is to find the optimal policy $\pi^*$ that maximizes the cumulative reward $\sum_{t=1}^{T} r_t$.
Policies must take into account the exploration versus exploitation dilemma and combine both explorative actions, to sample their associated unknown reward function to update their value-estimates, and greedy actions, to increase the total cumulative reward by choosing the action with the highest value-estimate. The main advantage of describing adaptive sampling in terms of a multi-armed bandit is that we can benefit from the extensive literature on bandits to find solutions and replace heuristic policies with more mathematically sound ones. 

\subsection{AdaptiveBandit}

Standard adaptive sampling algorithms work by performing several rounds or epochs of short parallel simulations. At each round, the algorithm is faced with the decision to select any of the sampled conformations from where to respawn a new round of simulations. The objective of these decisions is to avoid any redundant sampling and optimize our simulations to obtain the desired goal (which can be anything, from a full equilibrium characterization of a molecular system to sampling a specific conformation or dynamic event) at the minimum computational cost.

Here, we recast adaptive sampling in bandit terms, defining its tuple $\langle\mathcal{A},\mathcal{R},\gamma\rangle$. We define the action space $\mathcal{A}$ in terms of all possible conformations that are respawnable, i.e. they have been visited at least once,
\begin{equation}\label{samph}
    \mathcal{A}=\mathcal{H}_m = \{x_k \in \mathbb{R}^{3N}, k= 1, \dots, K_m\}\ ,
\end{equation}
where $K_m$ is the number of sampled configurations at epoch m.

There are different possible choices for the a priori unknown reward function $\mathcal{R}$ that the policy will try to maximize, and it will mostly depend on your objective with the simulation experiment. 

Because most of our MD experiments are usually aimed at sampling metastable states of interest, e.g. folded states of proteins or bound states between proteins and ligands, we have defined the reward $\mathcal{R}$ to be proportional to  minus equilibrium distribution so that that the optimal policy always picks conformations from the most stable state. Therefore, we define the reward $\mathcal{R}_a$ of action $a$  as the mean of the minus free energies of each configuration $x$ visited in the trajectory started with action $a$, i.e. 
\begin{equation}
    \mathcal{R}_a = <k_{B}T\log(\mu(x))>_{(a,x_1,\dots, x_{\tau})} \ ,
\end{equation}
where $\mu(x)$ is the equilibrium distribution and the average is computed over the succeeding frames in the trajectory starting from $a$.

The action space would be too large to compute meaningful value-estimations for each conformation, and there is no way to know the exact equilibrium distribution. To address this issue, we take advantage of MSM analysis to redefine the tuple $\langle\mathcal{A},\mathcal{R},\gamma\rangle$ in a more practical form. We define a reduced and tractable action space by using the MSM's discretized conformational space and use the stationary distribution of each state to obtain an estimate of their free energy to compute the rewards. We count each trajectory frame as an action taken, and use the succeeding frames to assign the reward. Because rewards strongly depend on how accurate the MSM estimation is, we use the latest MSM to recompute all past rewards from all trajectories at each epoch, differently from common Q-learning approaches \cite{sutton2018reinforcement}. Not only it ensures the best free energy estimation possible, but it also addresses the increasing action space problem, due to new conformations being sampled. Every epoch, the discretized conformational space is redefined, all frames are reassigned and rewards are recomputed on the newly defined states. 

\subsection{Solving the multi-armed bandit problem}

With the bandit tuple defined, we now need to deal with the exploration-exploitation trade-off and optimally solve it. To do so, \UCB\ relies on the UCB1 algorithm \cite{auer2002using} to optimize the action-picking policy, which defines the upper confidence bound for action values based on the number of times the agent has picked that action and the total number of actions taken. Therefore, actions are selected based on
\begin{equation}\label{equcb}
    a_t = \argmax_{a\in\mathcal{A}}\left[{Q_t(a) + c\sqrt{\frac{\ln{t}}{N_t(a)}}}\right],
\end{equation}
where $t$ denotes the total number of actions taken, $Q_t(a)$ is the estimated action-value for action $a$, $N_t(a)$ is the number of times action $a$ has been selected (prior to time $t$) and $c$ is a parameter controlling the degree of exploration. UCB1 follows the principle of \textit{"optimism in face of uncertainty"}, prioritizing actions with uncertain value-estimations, even if those values are not the greatest. To select an action, UCB1 not only takes into account the estimated value of that action, but also the amount of uncertainty on such value. By doing so, the algorithm not only promotes action exploration but also prioritizes the exploration of the most promising ones. In the long term, as our knowledge of action-values increases, the exploration term will decrease, and more greedy actions will be selected. UCB1 has a theoretical bound of $O(\sqrt{kTlog(L_t)})$ on its total regret $L_t$\cite{auer2002using}.

\subsection{\UCB\ with knowledge-based initialization\label{optinit}}

\UCB\ also has the option to initialize action-value estimates with external knowledge from the system, providing an initial value estimation to new actions, aiding to prioritize the most valuable actions. 
While in previous methods \cite{zimmerman2015fast,plattner2017complete} this is done by forcing the algorithm to sample from conformations based on a fixed empirical ranking, here we use the bandit formalism to initialize $Q$ in \Cref{equcb} with an empirical action-value function. This notably allows for the MSM to correct the initial prior suggestion for $Q$ given enough sampling. This is not true in previous schemes, where a partially wrong prior can affect sampling to the point of non-convergence to the intended results due to its degeneracy, i.e. even just some wrong contact information could kinetically bias the simulations far from the folding funnel. We demonstrate this aspect in the result section.  The initial prior $Q_{prior}(a)$ is computed as the average goal score from all frames in a state, and it is recalculated at each epoch, after re-clustering. The states are assigned with an initial pseudo-count $N_0(a)$, representing the statistical certainty of $Q_{prior}(a)$.

\subsection{Other adaptive sampling algorithms}

To evaluate \UCB's performance, we have tested it against several different adaptive sampling strategies, mainly the standard low-counts adaptive sampling, FAST \cite{zimmerman2015fast} and  Exploration-Exploitation.

The low-counts adaptive sampling is a simple and intuitive strategy that is optimal in pure exploration scenarios \cite{doerr2016htmd}. The method works by selecting conformations from the least populated clusters at each adaptive epoch. The other two methods, FAST and Exploration-Exploitation, are goal-oriented, were external knowledge on the system is used to guide sampling. 

FAST is also inspired by the multi-armed bandit problem, but the implementation differs as it uses an acquisition function to rank discrete conformational states rather than a reward function by definition, and actions (and their outcomes) do not influence their value-estimates. The acquisition function contains an exploitation term, defined by the goal scoring function that assigns a fixed value to each state, and an exploration term, based on state counts. The FAST implementation we used works as 
\begin{equation}
    \rho_i = \alpha\phi_i + (1-\alpha)\psi_i \ ,
\end{equation}
where $\rho_i$ is the score for state $i$, $\phi$ is the exploitative value obtained from the goal function for state $i$, $\psi$ is the exploration value defined by state $i$ counts (as in low-counts adaptive sampling) and $\alpha$ is a parameter regulating the weight of both terms. Both $\phi$ and $\psi$ terms are scaled to values that range from 0 to 1. The states are defined as the microstates obtained by the constructed Markov model at each epoch.

Lastly, we have Exploration-Exploitation, a strategy inspired by the popular method for multi-armed bandits $\epsilon$-greedy, implemented in HTMD's AdaptiveGoalEG \cite{doerr2016htmd}. Simulations are restarted $\epsilon$ times from the top goal ranking states, and $1-\epsilon$ times from the least sampled states (i.e the low-counts strategy). 

\subsection{Langevin dynamics on 2D Potentials}
We designed a set of experiments in a simple simulation set-up, performing Langevin dynamics on a single point mass of 1000 amu and a diffusion coefficient of 10{\AA}\textsuperscript{2}/ns at 300 K on two different potentials: a 2-wells potential (Figure \ref{fig:f1}a) inspired from Ref.~\citenum{pan2008building}, given by 
\begin{align}
   U(x,y) = &-3 e^{-(x-1)^2 - y^2} - 3 e^{-(x+1)^2 - y^2} 
   + 15e^{-0.32(x^2 + y^2 + 20(x+y)^2)}\\ \nonumber
   &+ 0.0512 (x^4+y^4) + 0.4 e^{-2-4y}
\end{align}
and a funnel potential (Figure \ref{fig:f1}c) given by 
\begin{align}
U(x,y)=2 \cos(2 \sqrt{x^2+y^2})-8 e^{-(x^2+y^2)}+0.2 ((x/8)^2+(y/8)^2)^3    
\end{align}
A reference baseline for each 2D potential was calculated using an MSM built with 10$\mu$s and 500$\mu$s  of aggregate simulation time for the 2-wells and funnel potential respectively, spawning trajectories from conformations covering the whole surface. Equilibrium probability was determined to be 50\% and 85\% respectively on each global minima. 

A total of 1 $\mu$s were simulated for each combination of method and potential, spawning 25 trajectories of 0.1 ns at each epoch for a total of 400 epochs. Performance at each epoch was measured as the mean of the equilibrium probabilities for the macrostate containing the targeted minimum for 10 independent MSMs built with 80\% of bootstrapped data. All the MSMs calculations were performed using HTMD \cite{doerr2016htmd}.

For the goal methods, we simulated a total of 2 $\mu$s for each method, spawning 10 trajectories per epoch with trajectories of 0.05 ns. Values of $\alpha = 0.1$ for FAST and $\epsilon = 0.1$ for Exploration-Exploitation were selected. In \UCB\, the exploration rate was set to $c=0.01$ and the initial pseudo-counts to $N_0(a) = 50$. 

\subsection{MD simulation set-up}
Simulation system for the chicken villin headpiece (PDB:2F4K) was built with HTMD \cite{doerr2016htmd}. We solvated villin in a 64\AA\  cubic box with a NaCl concentration of 0.05 $M$. Starting unfolded conformations for the runs were selected from a villin unfolding trajectory at high temperature (500 K).

In this context, we tested \UCB\, with $c=0.01$ and $N_0(a)=100$, against two different FAST setups, $\alpha = 0.5$ and $\alpha = 0.1$. A goal scoring function was used to guide the algorithms, based on the number of native C$\alpha$ contacts formed. For each setup, we ran parallel simulations of 10 ns, with 5 to 10 simulations per epoch, until we reached a total aggregate time of 4 $\mu$s. All simulations were run with ACEMD \cite{harvey2009acemd}, using the CHARMM22* force-field \cite{piana2011robust} on a local GPU cluster. A short HTMD code listing is provided as an example to run \UCB for villin simulations (Listing \ref{lst:code}).

\begin{lstlisting}[language=Python, label={lst:code},caption=Example AdaptiveBandit code, frame=tb, float,floatplacement=H]
from htmd.ui import *
from sklearn.cluster import MiniBatchKMeans
from jobqueues.localqueue import LocalGPUQueue
from goals import goalFunction

refmol = Molecule('villin_2f4k.pdb')
md = AdaptiveBandit()
md.app = LocalGPUQueue()
md.generatorspath = './generators'

md.clustmethod = MiniBatchKMeans
md.projection = MetricSelfDistance('protein and name CA')
md.goalfunction = delayed(goalFunction)(refmol)
md.ticadim = 3
md.nmin=5
md.nmax=10
md.nframes = 1000000

md.exploration = 0.01  ## "c" value
md.goal_init = 100     ## prior initialization value

md.run()

\end{lstlisting}

\section{Results}
\subsection{Performance testing on 2D Potentials}
\begin{figure}[!ht]
\centering
\includegraphics[width=\linewidth, keepaspectratio]{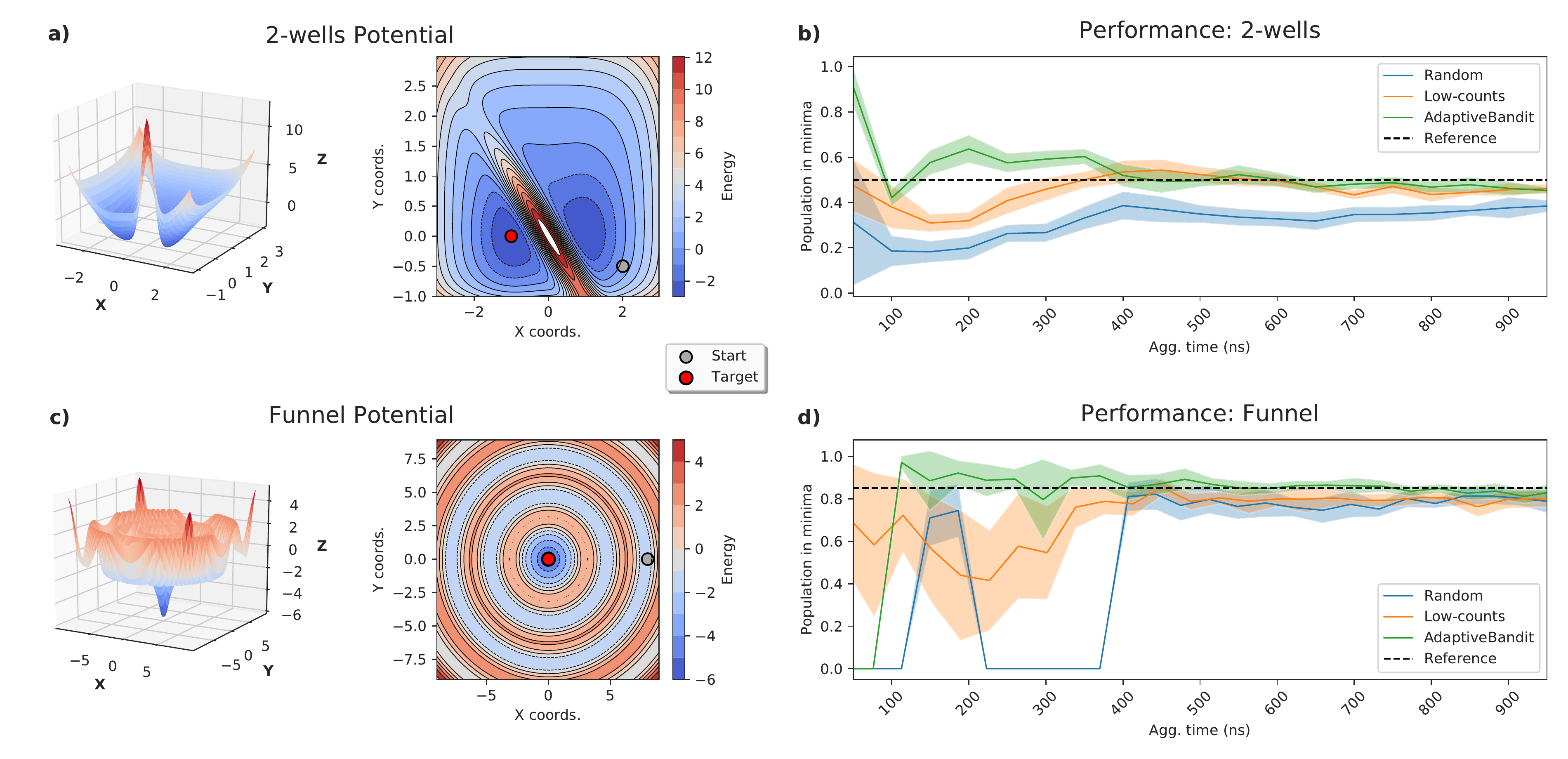}

\caption{\textbf{Performance comparison between random, low-counts and AdaptiveBandit in the experiments with 2D potentials.} 
\textbf{a), c)} 3D view and top view of the 2-wells and funnel potentials. Global minima are located at (-1, 0) and (0, 0) coordinates, respectively. Blue dot indicates starting points for simulations and red dot indicates the target global minima where population is measured at every epoch.
\textbf{b), d)} Performance comparison of total aggregate simulation time needed for random, low-counts and AdaptiveBandit sampling methods in 2-wells and funnel potential, respectively, to achieve correct  population estimates at their global minimum.
}

\label{fig:f1}
    
\end{figure}

The initial objective is to compare the performance of a set of adaptive sampling algorithms in a simple environment defined by 2D potentials. For this purpose, we performed Langevin dynamics on two different potentials: the 2-wells potential, composed of two minima separated by a high energetic barrier (Figure \ref{fig:f1}a), and a funnel potential, comprised of concentric isoenergetic regions with the global minimum located at its center (Figure \ref{fig:f1}c). The funnel potential is a useful benchmark to test the exploration-exploitation balance, as a purely exploratory strategy would tend to guide towards the outer circular wells, while the minimum is in the center. The objective of these experiments is to predict the equilibrium population of the targeted minima. The equilibrium populations are computed with MSM analysis to assess how different sampling strategies affect the MSM estimation.

First, \UCB\ was compared with two other common sampling policies, based on simple heuristics: random selection and the low-counts policy. Results for the 2-wells potential (Figure \ref{fig:f1}b) show a similar performance for the low-counts policy and \UCB. Both converge at the baseline population (50\%) while random sampling underestimates it. Because the potential just contains two large minima, comprising almost the entire conformational space, a fully explorative heuristic algorithm like the low-counts is optimal, as there is no need to prioritize anything besides exploring the two minima. \UCB\ is able to reach the same optimal performance.

\begin{figure}[!ht]
\centering
\includegraphics[width=\linewidth, keepaspectratio]{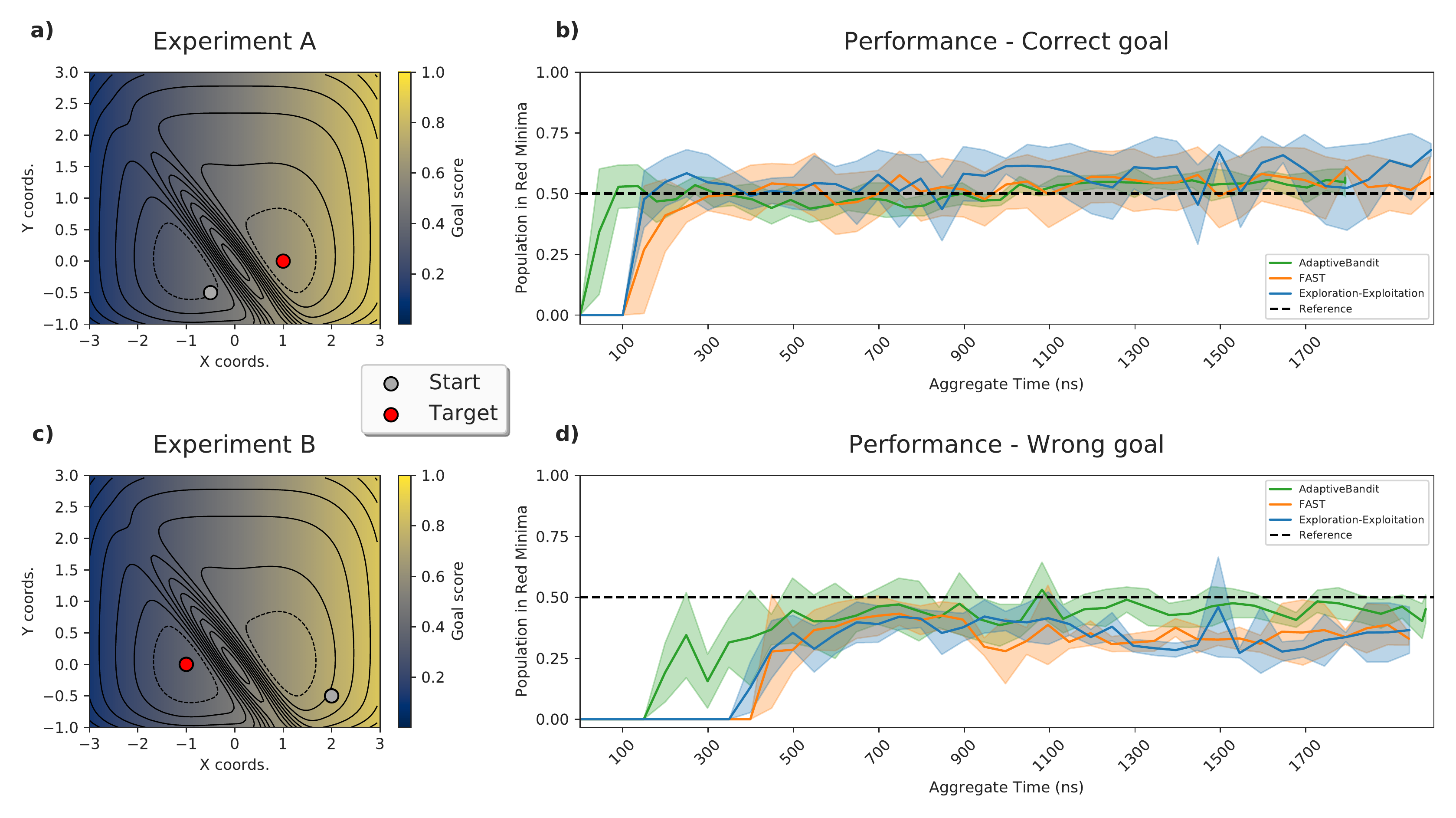}

\caption{\textbf{Performance comparison between goal-oriented algorithms FAST, $\textbf{epsilon}$-greedy and AdaptiveBandit in the experiments with 2D potentials.}  
\textbf{a), c)} Top view of 2-wells potential. Goal distribution across the potential is shown. Blue dots indicate the starting conformations for the runs. Red dots indicate the minima where population is measured.
\textbf{b), d)} Performance comparison of total aggregate simulation time needed for  FAST, Exploration-Exploitation and AdaptiveBandit  methods to correctly estimate populations at their target minimum.
}
\label{fig:f2}
    
\end{figure}

\begin{figure}[!ht]
\centering
\includegraphics[width=\linewidth, keepaspectratio]{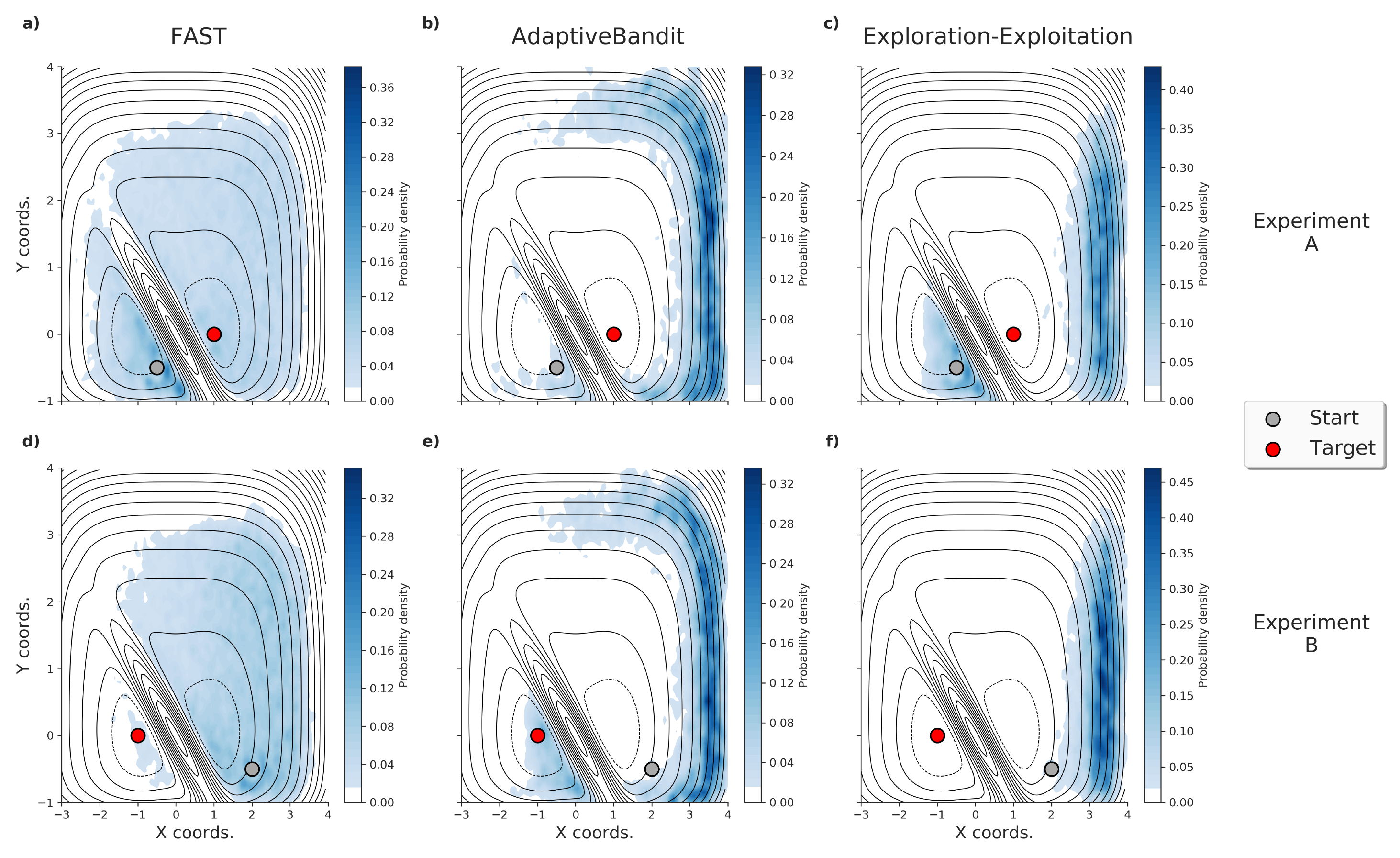}

\caption{\textbf{Simulation re-spawning distribution by algorithm across the 2-wells potential.} 
Each plot depicts the probability distribution of selected conformations throughout the runs, obtained by kernel density estimate \cite{kdeplot}. Starting points for each run are represented with a blue dot and target minimum with a red one. Goal distribution (not shown) is the same as in \Cref{fig:f2}. Subplots \textbf{a), b), c)} represent the spawning probability distribution across the potential surface on experiment A, target minima at coordinates (1, 0), for FAST, AdaptiveBandit and Exploration-Exploitation algorithms, and \textbf{d), e), f)} for experiment B, target minima at coordinates (0, -0.5). 
}
\label{fig:f3}
    
\end{figure}

For the funnel potential (Figure \ref{fig:f1}d), the relative size of the minima is much smaller compared to the conformational space, hence its detection by random sampling is more inefficient than for the other two algorithms. The low-counts method is able to reach the minima faster, as it is to cover the space quickly. Both these algorithms obtain a slight underestimation of the equilibrium population. On the other hand, \UCB\ achieves a more accurate estimation and reaches convergence with 4 times less aggregate time than the other algorithms, highly reducing the computational resources needed to obtain accurate estimations of the equilibrium distribution. 

This first test here showcases how introducing an exploitation term to quantify an action's value, besides the exploration term, either increases or equals the performance of fully exploratory algorithms on obtaining correct equilibrium estimations in the tested systems. Value-estimations of each action help on prioritizing sampling on the most relevant areas of the conformational space, rather than just exploring everything and sampling irrelevant conformations. While in the 2-wells potential this does not make a big difference, it does in the funnel potential, where \UCB\ focuses sampling on the minima by identifying its relevance with action value-estimates and does not waste resources on exploring irrelevant conformations.

\subsection{Using system external knowledge}
Next, we want to test how \UCB\ performs in the 2-wells potential against two existing methods that incorporate an exploitative term by employing external knowledge on the system. The pair of tested algorithms, also known as goal-oriented methods, are FAST\cite{zimmerman2015fast} and Exploration-Exploitation. To make sure \UCB\ is at the same level of system knowledge as the other methods, the information provided by the goal-function was used in \UCB\ through knowledge-based initialization (as explained in \textit{Methods}).

The goal function employed in the experiments with the 2-wells potential increases the score linearly with the $x$ axis (\Cref{fig:f2}a,c), thus creating a gradient of reward pushing to the right boundary of the potential. Two tests were performed in different scenarios. In the first test, the target minimum has a greater score than the starting coordinate (Experiment A, Figure \ref{fig:f2}a). In the second one, the target minimum has a lower goal than the starting conformations, and therefore requires opposition to the goal's influence to obtain accurate estimations on the target minimum (Experiment B, Figure \ref{fig:f2}c).

For experiment A, all methods reached the reference population, with \UCB\ needing slightly less simulation time to reach the correct population estimation in the target (\ref{fig:f2}b). Differences in the algorithms can be visualized by a distribution plot of the spawning conformations in Figure \ref{fig:f3}. 
During the initial epochs, both FAST and Exploration-Exploitation follow the goal, spawning new simulations pushing against the energy barrier. \UCB, on the other hand, quickly discovers the target minima and starts exploring other areas and not only directs sampling on the high score region but also in its surroundings. Even though the performance of all three algorithms is similar, differences in the spawning patterns between the three algorithms can be appreciated throughout the experiment. FAST presents a more explorative behavior and respawns simulations from all along the conformational space (Figure \ref{fig:f3}a). On the other hand, Exploration-Exploitation presents a highly exploitative behavior, strongly focusing on the highest goal-scoring region once it is discovered (Figure \ref{fig:f3}c). In between, \UCB\ presents an overall greedy behavior, but with higher levels of exploration than the Exploration-Exploitation method which translates into a small boost in its performance. It is interesting to point out the few resources invested by \UCB in the origin minima, which demonstrates that the algorithm quickly identifies it as a non-interesting area (Figure \ref{fig:f3}b).

For experiment B, \UCB\ reaches the target minimum faster and equilibrium populations are estimated more accurately (Figure \ref{fig:f2}d). Both Exploration-Exploitation and FAST require more simulation time to reach the target minima and fail to converge on the correct equilibrium populations. In this scenario, Exploration-Exploitation is greatly focused in the high scoring region (\Cref{fig:f3}f) resulting in a marginal exploration of the target minimum, while FAST and specially \UCB\ do perform a more significant search on it (\Cref{fig:f3}d,e). Comparison between \UCB\ and FAST spawning patterns (\Cref{fig:f3}d,e) reveals the differences in the exploration profile, where again FAST thoroughly spawns conformations from every explored point in the surface, while \UCB , following the goal, explores the boundaries of the conformational space. Even if differences in performance are not substantially large, the experiment shows us the inability of FAST and Exploration-Exploitation to update the initial action-value estimates, translating into a lack of adaptation to the system being sampled. In opposition, \UCB\ is able to correct the prior action-value estimates and readjust the sampling policy to a more optimal one, as it uses exploitation intrinsically based on MSM estimations from the available simulation data and external knowledge is introduced as prior information, rather than as the function to optimize.  The ability to update the system knowledge at each epoch is crucial in experiments where the goal scoring function used has high levels of degeneracy or is directly wrong. Asymptotically, \UCB\  should always be better as it is logarithmically bound on the number of trials to the total regret \cite{auer2002using} (the difference between the maximum possible reward and the current reward), whereas Exploration-Exploitation and FAST are linearly bound.

\subsection{Testing on protein folding simulations}
\begin{figure}[!ht]
\centering
\includegraphics[width=\linewidth, keepaspectratio]{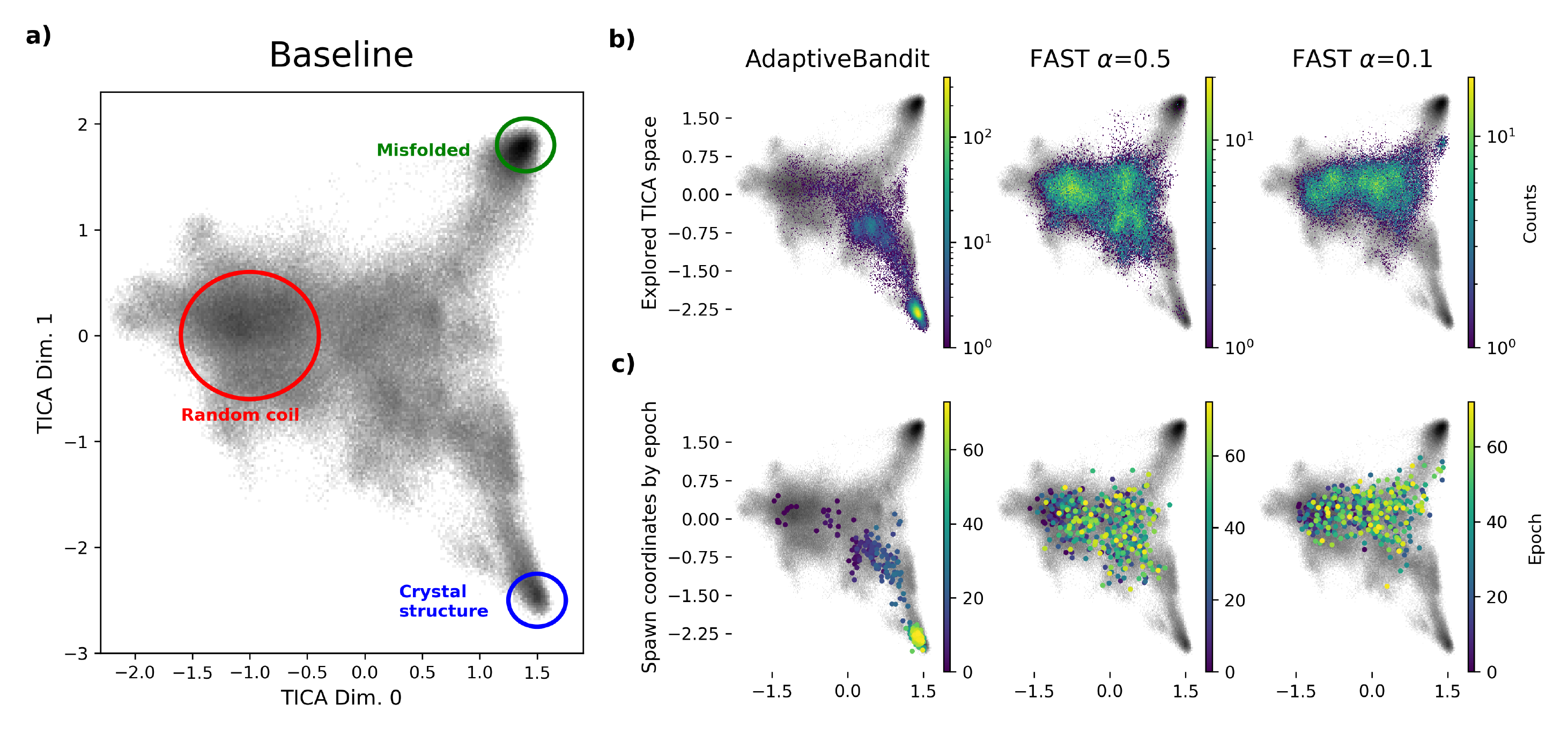}

\caption{\textbf{Villin folding simulations.} 
\textbf{a)} Conformational space for folding of villin on the baseline data set. The tICA space includes large regions of random coil (initial conformation are located within the
red circle), misfolded conformations (green circle), and crystal-like structures (blue circle). 
\textbf{b)} Exploration of the conformational space by sampling algorithms. Each plot includes the baseline exploration depicted on gray and the explored space with a colored heatmap.
\textbf{c)} Spamming coordinates for new epochs. Scattered points indicate starting conformations for new epochs, colored from first (purple) to last (yellow).
}
\label{fig:f4}
    
\end{figure}

Besides testing in simple 2D potentials, we explored \UCB's performance on a more realistic and challenging scenario. \UCB\ was tested on protein folding simulations, using villin as a benchmark. The chicken villin headpiece consists of a chain of 35 residues that folds into a three $\alpha$-helical bundle, sharing a common hydrophobic core \cite{kubelka2006sub}. It is known to have a fast folding rate of (0.7 $\mu$s)\textsuperscript{-1} \cite{kubelka2006sub}. Our target for this test is to reach the folded state with the minimum amount of aggregate time and compare how \UCB\  and FAST distribute sampling across the conformational space of villin. Because we are testing the algorithm's effect rather than the technical capabilities of reaching villin's folding state with MD, we set up very short simulation times to increase the number of epochs and ensure we are evaluating the algorithm's performance. The goal function used for the algorithms maximizes the number of native C$\alpha$ contacts formed to guide sampling on to the folded state. 30 $\mu$s of villin folding simulations were used to build some reference tICA dimensions to evaluate the sampled conformational space from each method. The first two TICA dimensions reveal three main states (Figure \ref{fig:f4}a): the unfolded state (random coil), the folded structure, and a misfolded state.

Figure \ref{fig:f4}b shows the distinctive behavior of \UCB\ and FAST while sampling the folding path. \UCB\ clearly reaches the crystal structure. FAST struggles to do so due to the very short trajectories used, which produces a sampling bias, as indicated in \cite{wan2020adaptive}. The results showcase how \UCB\ is able to select the most relevant conformations to reach the folded state, prioritizing the most promising actions from the subset of undersampled actions. On the contrary, FAST, even in its most greedy setting ($\alpha$ = 0.1), is not able to correctly prioritize the most relevant states and keeps exploring over random coil states, even in the latest epochs (Figure \ref{fig:f4}c). 
The greedy setting also presents a slight misdirection towards the misfolded state, which suggests that the used goal scoring function has degeneracy and it does not differentiate enough between native-like structures and misfolded structures that are very far dynamically. As commented in the previous experiment using external knowledge on the 2-wells potential, methods like FAST or Exploration-Exploitation that rely only on external information can be severely hampered when the provided information does not represent the true energetic gradient. \UCB\ prevents that by updating the prior information with rewards coming from interacting with the system and observing its response to our actions.

\section{Conclusion}

\UCB\ formally introduces adaptive sampling into reinforcement learning by describing it in terms of multi-armed bandits and builds upon it to deliver a novel algorithm with increased performance and flexibility across different energy landscapes. 
\UCB\ is able to perform equally or better than previous adaptive sampling algorithms in a  diverse set of systems, and it has demonstrated the ability to learn from simulation results. \UCB\ works both with and without external knowledge of the system, and it can update prior beliefs in the system based on the results obtained during the experiment.

Goal-oriented adaptive sampling methods as in Ref. \citenum{zimmerman2015fast} also get inspiration from exploration-exploitation strategies, like $epsilon$-greedy.
The context, however, is quite different as there is not a definition of a multi-armed bandit framework and a reward per action, rather it is more akin to directly define an acquisition function. 
Furthermore, the greediness is towards predetermined states given from external knowledge on the system. 
AdaptiveBandit, as used here, uses exploitation intrinsically without requiring external information. 
It is, however, a possibility to do so and use experimental data to provide a prior for the sampling.

We have exemplified here cases were \UCB\ works better due to its adaptability and flexibility, but that does not mean that it could underperform in other scenarios. Our implementation of \UCB\ relies on good MSM estimates, and therefore the action-value estimates carry on with errors caused not only by discretization and dimensionality reduction but also by the sampling bias, especially on estimations of equilibrium populations \cite{wan2020adaptive}. Additionally, \UCB's performance also depends on the $c$ hyperparameter to regulate exploration and it is not very intuitive, as it must be tuned according to the scale of both terms in \Cref{equcb}.

The version of \UCB\ presented here defines a reward proportional to the free energy of each state and utilizes the UCB1 algorithm to optimize the action-picking policy. However, this is not the only possible way to apply \UCB\, and the algorithm can be changed to better adapt the experiment and systems. We hope that our work inspires the development of new adaptive sampling algorithms built under theoretical fundamentals instead of using simple heuristic policies.

\begin{acknowledgement}

G.D.F. acknowledges support from MINECO (Unidad de Excelencia Mar{\'i}a de Maeztu MDM-2014-0370 and BIO2017-82628-P) FEDER and Secretaria d'Universitats i Recerca de la Generalitat de Catalunya. This project received funding from the European Union's Horizon 2020 Research and Innovation Programme under Grant Agreement 675451 (CompBioMed Project).
\end{acknowledgement}


\bibliography{main}

\providecommand{\latin}[1]{#1}
\makeatletter
\providecommand{\doi}
  {\begingroup\let\do\@makeother\dospecials
  \catcode`\{=1 \catcode`\}=2 \doi@aux}
\providecommand{\doi@aux}[1]{\endgroup\texttt{#1}}
\makeatother
\providecommand*\mcitethebibliography{\thebibliography}
\csname @ifundefined\endcsname{endmcitethebibliography}
  {\let\endmcitethebibliography\endthebibliography}{}
\begin{mcitethebibliography}{47}
\providecommand*\natexlab[1]{#1}
\providecommand*\mciteSetBstSublistMode[1]{}
\providecommand*\mciteSetBstMaxWidthForm[2]{}
\providecommand*\mciteBstWouldAddEndPuncttrue
  {\def\EndOfBibitem{\unskip.}}
\providecommand*\mciteBstWouldAddEndPunctfalse
  {\let\EndOfBibitem\relax}
\providecommand*\mciteSetBstMidEndSepPunct[3]{}
\providecommand*\mciteSetBstSublistLabelBeginEnd[3]{}
\providecommand*\EndOfBibitem{}
\mciteSetBstSublistMode{f}
\mciteSetBstMaxWidthForm{subitem}{(\alph{mcitesubitemcount})}
\mciteSetBstSublistLabelBeginEnd
  {\mcitemaxwidthsubitemform\space}
  {\relax}
  {\relax}

\bibitem[Martinez-Rosell \latin{et~al.}(2017)Martinez-Rosell, Giorgino, Harvey,
  and de~Fabritiis]{martinez2017drug}
Martinez-Rosell,~G.; Giorgino,~T.; Harvey,~M.~J.; de~Fabritiis,~G.
  \emph{Current topics in medicinal chemistry} \textbf{2017}, \emph{17},
  2617--2625\relax
\mciteBstWouldAddEndPuncttrue
\mciteSetBstMidEndSepPunct{\mcitedefaultmidpunct}
{\mcitedefaultendpunct}{\mcitedefaultseppunct}\relax
\EndOfBibitem
\bibitem[P{\'e}rez \latin{et~al.}(2018)P{\'e}rez, Mart{\'\i}nez-Rosell, and
  De~Fabritiis]{perez2018simulations}
P{\'e}rez,~A.; Mart{\'\i}nez-Rosell,~G.; De~Fabritiis,~G. \emph{Current opinion
  in structural biology} \textbf{2018}, \emph{49}, 139--144\relax
\mciteBstWouldAddEndPuncttrue
\mciteSetBstMidEndSepPunct{\mcitedefaultmidpunct}
{\mcitedefaultendpunct}{\mcitedefaultseppunct}\relax
\EndOfBibitem
\bibitem[Torrie and Valleau(1977)Torrie, and Valleau]{torrie1977nonphysical}
Torrie,~G.~M.; Valleau,~J.~P. \emph{Journal of Computational Physics}
  \textbf{1977}, \emph{23}, 187--199\relax
\mciteBstWouldAddEndPuncttrue
\mciteSetBstMidEndSepPunct{\mcitedefaultmidpunct}
{\mcitedefaultendpunct}{\mcitedefaultseppunct}\relax
\EndOfBibitem
\bibitem[Frenkel and Smit(1996)Frenkel, and Smit]{Frenkel:1996:UMS:547952}
Frenkel,~D., Smit,~B., Eds. \emph{Understanding Molecular Simulation: From
  Algorithms to Applications}, 1st ed.; Academic Press, Inc.: Orlando, FL, USA,
  1996\relax
\mciteBstWouldAddEndPuncttrue
\mciteSetBstMidEndSepPunct{\mcitedefaultmidpunct}
{\mcitedefaultendpunct}{\mcitedefaultseppunct}\relax
\EndOfBibitem
\bibitem[Sugita and Okamoto(1999)Sugita, and Okamoto]{sugita1999replica}
Sugita,~Y.; Okamoto,~Y. \emph{Chemical physics letters} \textbf{1999},
  \emph{314}, 141--151\relax
\mciteBstWouldAddEndPuncttrue
\mciteSetBstMidEndSepPunct{\mcitedefaultmidpunct}
{\mcitedefaultendpunct}{\mcitedefaultseppunct}\relax
\EndOfBibitem
\bibitem[Fukunishi \latin{et~al.}(2002)Fukunishi, Watanabe, and
  Takada]{fukunishi2002hamiltonian}
Fukunishi,~H.; Watanabe,~O.; Takada,~S. \emph{The Journal of chemical physics}
  \textbf{2002}, \emph{116}, 9058--9067\relax
\mciteBstWouldAddEndPuncttrue
\mciteSetBstMidEndSepPunct{\mcitedefaultmidpunct}
{\mcitedefaultendpunct}{\mcitedefaultseppunct}\relax
\EndOfBibitem
\bibitem[Izrailev \latin{et~al.}(1999)Izrailev, Stepaniants, Isralewitz,
  Kosztin, Lu, Molnar, Wriggers, and Schulten]{izrailev1999steered}
Izrailev,~S.; Stepaniants,~S.; Isralewitz,~B.; Kosztin,~D.; Lu,~H.; Molnar,~F.;
  Wriggers,~W.; Schulten,~K. \emph{Computational molecular dynamics:
  challenges, methods, ideas}; Springer, 1999; pp 39--65\relax
\mciteBstWouldAddEndPuncttrue
\mciteSetBstMidEndSepPunct{\mcitedefaultmidpunct}
{\mcitedefaultendpunct}{\mcitedefaultseppunct}\relax
\EndOfBibitem
\bibitem[Isralewitz \latin{et~al.}(2001)Isralewitz, Gao, and
  Schulten]{isralewitz2001steered}
Isralewitz,~B.; Gao,~M.; Schulten,~K. \emph{Current opinion in structural
  biology} \textbf{2001}, \emph{11}, 224--230\relax
\mciteBstWouldAddEndPuncttrue
\mciteSetBstMidEndSepPunct{\mcitedefaultmidpunct}
{\mcitedefaultendpunct}{\mcitedefaultseppunct}\relax
\EndOfBibitem
\bibitem[Laio and Parrinello(2002)Laio, and Parrinello]{laio2002escaping}
Laio,~A.; Parrinello,~M. \emph{Proceedings of the National Academy of Sciences}
  \textbf{2002}, \emph{99}, 12562--12566\relax
\mciteBstWouldAddEndPuncttrue
\mciteSetBstMidEndSepPunct{\mcitedefaultmidpunct}
{\mcitedefaultendpunct}{\mcitedefaultseppunct}\relax
\EndOfBibitem
\bibitem[Rezende and Mohamed(2015)Rezende, and Mohamed]{rezende2015variational}
Rezende,~D.~J.; Mohamed,~S. \emph{arXiv preprint arXiv:1505.05770}
  \textbf{2015}, \relax
\mciteBstWouldAddEndPunctfalse
\mciteSetBstMidEndSepPunct{\mcitedefaultmidpunct}
{}{\mcitedefaultseppunct}\relax
\EndOfBibitem
\bibitem[No{\'e} \latin{et~al.}(2019)No{\'e}, Olsson, K{\"o}hler, and
  Wu]{noe2019boltzmann}
No{\'e},~F.; Olsson,~S.; K{\"o}hler,~J.; Wu,~H. \emph{Science} \textbf{2019},
  \emph{365}, eaaw1147\relax
\mciteBstWouldAddEndPuncttrue
\mciteSetBstMidEndSepPunct{\mcitedefaultmidpunct}
{\mcitedefaultendpunct}{\mcitedefaultseppunct}\relax
\EndOfBibitem
\bibitem[Singhal and Pande(2005)Singhal, and Pande]{singhal2005error}
Singhal,~N.; Pande,~V.~S. \emph{The Journal of chemical physics} \textbf{2005},
  \emph{123}, 204909\relax
\mciteBstWouldAddEndPuncttrue
\mciteSetBstMidEndSepPunct{\mcitedefaultmidpunct}
{\mcitedefaultendpunct}{\mcitedefaultseppunct}\relax
\EndOfBibitem
\bibitem[Hinrichs and Pande(2007)Hinrichs, and Pande]{hinrichs2007calculation}
Hinrichs,~N.~S.; Pande,~V.~S. \emph{The Journal of chemical physics}
  \textbf{2007}, \emph{126}, 244101\relax
\mciteBstWouldAddEndPuncttrue
\mciteSetBstMidEndSepPunct{\mcitedefaultmidpunct}
{\mcitedefaultendpunct}{\mcitedefaultseppunct}\relax
\EndOfBibitem
\bibitem[Pronk \latin{et~al.}(2011)Pronk, Larsson, Pouya, Bowman, Haque,
  Beauchamp, Hess, Pande, Kasson, and Lindahl]{pronk2011copernicus}
Pronk,~S.; Larsson,~P.; Pouya,~I.; Bowman,~G.~R.; Haque,~I.~S.; Beauchamp,~K.;
  Hess,~B.; Pande,~V.~S.; Kasson,~P.~M.; Lindahl,~E. Copernicus: A new paradigm
  for parallel adaptive molecular dynamics. Proceedings of 2011 International
  Conference for High Performance Computing, Networking, Storage and Analysis.
  2011; p~60\relax
\mciteBstWouldAddEndPuncttrue
\mciteSetBstMidEndSepPunct{\mcitedefaultmidpunct}
{\mcitedefaultendpunct}{\mcitedefaultseppunct}\relax
\EndOfBibitem
\bibitem[Doerr and De~Fabritiis(2014)Doerr, and De~Fabritiis]{doerr2014fly}
Doerr,~S.; De~Fabritiis,~G. \emph{Journal of chemical theory and computation}
  \textbf{2014}, \emph{10}, 2064--2069\relax
\mciteBstWouldAddEndPuncttrue
\mciteSetBstMidEndSepPunct{\mcitedefaultmidpunct}
{\mcitedefaultendpunct}{\mcitedefaultseppunct}\relax
\EndOfBibitem
\bibitem[No{\'e} \latin{et~al.}(2009)No{\'e}, Sch{\"u}tte, Vanden-Eijnden,
  Reich, and Weikl]{noe2009constructing}
No{\'e},~F.; Sch{\"u}tte,~C.; Vanden-Eijnden,~E.; Reich,~L.; Weikl,~T.~R.
  \emph{Proceedings of the National Academy of Sciences} \textbf{2009},
  \emph{106}, 19011--19016\relax
\mciteBstWouldAddEndPuncttrue
\mciteSetBstMidEndSepPunct{\mcitedefaultmidpunct}
{\mcitedefaultendpunct}{\mcitedefaultseppunct}\relax
\EndOfBibitem
\bibitem[Plattner \latin{et~al.}(2017)Plattner, Doerr, De~Fabritiis, and
  No{\'e}]{plattner2017complete}
Plattner,~N.; Doerr,~S.; De~Fabritiis,~G.; No{\'e},~F. \emph{Nature chemistry}
  \textbf{2017}, \emph{9}, 1005\relax
\mciteBstWouldAddEndPuncttrue
\mciteSetBstMidEndSepPunct{\mcitedefaultmidpunct}
{\mcitedefaultendpunct}{\mcitedefaultseppunct}\relax
\EndOfBibitem
\bibitem[Shaw \latin{et~al.}(2008)Shaw, Deneroff, Dror, Kuskin, Larson, Salmon,
  Young, Batson, Bowers, Chao, and et~al.]{shaw2008anton}
Shaw,~D.~E.; Deneroff,~M.~M.; Dror,~R.~O.; Kuskin,~J.~S.; Larson,~R.~H.;
  Salmon,~J.~K.; Young,~C.; Batson,~B.; Bowers,~K.~J.; Chao,~J.~C.; et~al.,
  \emph{Commun. ACM} \textbf{2008}, \emph{51}, 91--97\relax
\mciteBstWouldAddEndPuncttrue
\mciteSetBstMidEndSepPunct{\mcitedefaultmidpunct}
{\mcitedefaultendpunct}{\mcitedefaultseppunct}\relax
\EndOfBibitem
\bibitem[Lindorff-Larsen \latin{et~al.}(2011)Lindorff-Larsen, Piana, Dror, and
  Shaw]{lindorff2011fast}
Lindorff-Larsen,~K.; Piana,~S.; Dror,~R.~O.; Shaw,~D.~E. \emph{Science}
  \textbf{2011}, \emph{334}, 517--520\relax
\mciteBstWouldAddEndPuncttrue
\mciteSetBstMidEndSepPunct{\mcitedefaultmidpunct}
{\mcitedefaultendpunct}{\mcitedefaultseppunct}\relax
\EndOfBibitem
\bibitem[Friedrichs \latin{et~al.}(2009)Friedrichs, Eastman, Vaidyanathan,
  Houston, Legrand, Beberg, Ensign, Bruns, and
  Pande]{friedrichs2009accelerating}
Friedrichs,~M.~S.; Eastman,~P.; Vaidyanathan,~V.; Houston,~M.; Legrand,~S.;
  Beberg,~A.~L.; Ensign,~D.~L.; Bruns,~C.~M.; Pande,~V.~S. \emph{Journal of
  computational chemistry} \textbf{2009}, \emph{30}, 864--872\relax
\mciteBstWouldAddEndPuncttrue
\mciteSetBstMidEndSepPunct{\mcitedefaultmidpunct}
{\mcitedefaultendpunct}{\mcitedefaultseppunct}\relax
\EndOfBibitem
\bibitem[Harvey and De~Fabritiis(2009)Harvey, and
  De~Fabritiis]{harvey2009implementation}
Harvey,~M.; De~Fabritiis,~G. \emph{Journal of chemical theory and computation}
  \textbf{2009}, \emph{5}, 2371--2377\relax
\mciteBstWouldAddEndPuncttrue
\mciteSetBstMidEndSepPunct{\mcitedefaultmidpunct}
{\mcitedefaultendpunct}{\mcitedefaultseppunct}\relax
\EndOfBibitem
\bibitem[Harvey \latin{et~al.}(2009)Harvey, Giupponi, and
  Fabritiis]{harvey2009acemd}
Harvey,~M.~J.; Giupponi,~G.; Fabritiis,~G.~D. \emph{Journal of chemical theory
  and computation} \textbf{2009}, \emph{5}, 1632--1639\relax
\mciteBstWouldAddEndPuncttrue
\mciteSetBstMidEndSepPunct{\mcitedefaultmidpunct}
{\mcitedefaultendpunct}{\mcitedefaultseppunct}\relax
\EndOfBibitem
\bibitem[Eastman \latin{et~al.}(2017)Eastman, Swails, Chodera, McGibbon, Zhao,
  Beauchamp, Wang, Simmonett, Harrigan, Stern, Wiewiora, Brooks, and
  Pande]{Eastman2017}
Eastman,~P.; Swails,~J.; Chodera,~J.~D.; McGibbon,~R.~T.; Zhao,~Y.;
  Beauchamp,~K.~A.; Wang,~L.~P.; Simmonett,~A.~C.; Harrigan,~M.~P.;
  Stern,~C.~D.; Wiewiora,~R.~P.; Brooks,~B.~R.; Pande,~V.~S. \emph{PLoS Comput.
  Biol.} \textbf{2017}, \emph{13}\relax
\mciteBstWouldAddEndPuncttrue
\mciteSetBstMidEndSepPunct{\mcitedefaultmidpunct}
{\mcitedefaultendpunct}{\mcitedefaultseppunct}\relax
\EndOfBibitem
\bibitem[Prinz \latin{et~al.}(2011)Prinz, Wu, Sarich, Keller, Senne, Held,
  Chodera, Sch{\"u}tte, and No{\'e}]{prinz2011markov}
Prinz,~J.-H.; Wu,~H.; Sarich,~M.; Keller,~B.; Senne,~M.; Held,~M.;
  Chodera,~J.~D.; Sch{\"u}tte,~C.; No{\'e},~F. \emph{The Journal of chemical
  physics} \textbf{2011}, \emph{134}, 174105\relax
\mciteBstWouldAddEndPuncttrue
\mciteSetBstMidEndSepPunct{\mcitedefaultmidpunct}
{\mcitedefaultendpunct}{\mcitedefaultseppunct}\relax
\EndOfBibitem
\bibitem[Bowman \latin{et~al.}(2013)Bowman, Pande, and
  No{\'e}]{bowman2013introduction}
Bowman,~G.~R.; Pande,~V.~S.; No{\'e},~F. \emph{An introduction to Markov state
  models and their application to long timescale molecular simulation};
  Springer Science \& Business Media, 2013; Vol. 797\relax
\mciteBstWouldAddEndPuncttrue
\mciteSetBstMidEndSepPunct{\mcitedefaultmidpunct}
{\mcitedefaultendpunct}{\mcitedefaultseppunct}\relax
\EndOfBibitem
\bibitem[Buch \latin{et~al.}(2011)Buch, Giorgino, and
  De~Fabritiis]{buch2011complete}
Buch,~I.; Giorgino,~T.; De~Fabritiis,~G. \emph{Proceedings of the National
  Academy of Sciences} \textbf{2011}, \emph{108}, 10184--10189\relax
\mciteBstWouldAddEndPuncttrue
\mciteSetBstMidEndSepPunct{\mcitedefaultmidpunct}
{\mcitedefaultendpunct}{\mcitedefaultseppunct}\relax
\EndOfBibitem
\bibitem[Mart{\'\i}nez-Rosell \latin{et~al.}(2018)Mart{\'\i}nez-Rosell, Harvey,
  and De~Fabritiis]{martinez2018molecular}
Mart{\'\i}nez-Rosell,~G.; Harvey,~M.~J.; De~Fabritiis,~G. \emph{Journal of
  chemical information and modeling} \textbf{2018}, \emph{58}, 683--691\relax
\mciteBstWouldAddEndPuncttrue
\mciteSetBstMidEndSepPunct{\mcitedefaultmidpunct}
{\mcitedefaultendpunct}{\mcitedefaultseppunct}\relax
\EndOfBibitem
\bibitem[Sabbadin and Moro(2014)Sabbadin, and Moro]{sabbadin2014supervised}
Sabbadin,~D.; Moro,~S. \emph{Journal of chemical information and modeling}
  \textbf{2014}, \emph{54}, 372--376\relax
\mciteBstWouldAddEndPuncttrue
\mciteSetBstMidEndSepPunct{\mcitedefaultmidpunct}
{\mcitedefaultendpunct}{\mcitedefaultseppunct}\relax
\EndOfBibitem
\bibitem[Perez \latin{et~al.}(2015)Perez, MacCallum, and
  Dill]{perez2015accelerating}
Perez,~A.; MacCallum,~J.~L.; Dill,~K.~A. \emph{Proceedings of the National
  Academy of Sciences} \textbf{2015}, \emph{112}, 11846--11851\relax
\mciteBstWouldAddEndPuncttrue
\mciteSetBstMidEndSepPunct{\mcitedefaultmidpunct}
{\mcitedefaultendpunct}{\mcitedefaultseppunct}\relax
\EndOfBibitem
\bibitem[Zimmerman and Bowman(2015)Zimmerman, and Bowman]{zimmerman2015fast}
Zimmerman,~M.~I.; Bowman,~G.~R. \emph{Journal of chemical theory and
  computation} \textbf{2015}, \emph{11}, 5747--5757\relax
\mciteBstWouldAddEndPuncttrue
\mciteSetBstMidEndSepPunct{\mcitedefaultmidpunct}
{\mcitedefaultendpunct}{\mcitedefaultseppunct}\relax
\EndOfBibitem
\bibitem[Ovchinnikov \latin{et~al.}(2017)Ovchinnikov, Park, Varghese, Huang,
  Pavlopoulos, Kim, Kamisetty, Kyrpides, and Baker]{ovchinnikov2017protein}
Ovchinnikov,~S.; Park,~H.; Varghese,~N.; Huang,~P.-S.; Pavlopoulos,~G.~A.;
  Kim,~D.~E.; Kamisetty,~H.; Kyrpides,~N.~C.; Baker,~D. \emph{Science}
  \textbf{2017}, \emph{355}, 294--298\relax
\mciteBstWouldAddEndPuncttrue
\mciteSetBstMidEndSepPunct{\mcitedefaultmidpunct}
{\mcitedefaultendpunct}{\mcitedefaultseppunct}\relax
\EndOfBibitem
\bibitem[Zimmerman \latin{et~al.}(2017)Zimmerman, Hart, Sibbald, Frederick,
  Jimah, Knoverek, Tolia, and Bowman]{zimmerman2017prediction}
Zimmerman,~M.~I.; Hart,~K.~M.; Sibbald,~C.~A.; Frederick,~T.~E.; Jimah,~J.~R.;
  Knoverek,~C.~R.; Tolia,~N.~H.; Bowman,~G.~R. \emph{ACS central science}
  \textbf{2017}, \emph{3}, 1311--1321\relax
\mciteBstWouldAddEndPuncttrue
\mciteSetBstMidEndSepPunct{\mcitedefaultmidpunct}
{\mcitedefaultendpunct}{\mcitedefaultseppunct}\relax
\EndOfBibitem
\bibitem[Cruz \latin{et~al.}(2020)Cruz, Frederick, Singh, Vithani, Zimmerman,
  Porter, Moeder, Amarasinghe, and Bowman]{cruz2020discovery}
Cruz,~M.~A.; Frederick,~T.~E.; Singh,~S.; Vithani,~N.; Zimmerman,~M.~I.;
  Porter,~J.~R.; Moeder,~K.~E.; Amarasinghe,~G.~K.; Bowman,~G.~R.
  \emph{bioRxiv} \textbf{2020}, \relax
\mciteBstWouldAddEndPunctfalse
\mciteSetBstMidEndSepPunct{\mcitedefaultmidpunct}
{}{\mcitedefaultseppunct}\relax
\EndOfBibitem
\bibitem[Zimmerman \latin{et~al.}(2018)Zimmerman, Porter, Sun, Silva, and
  Bowman]{zimmerman2018choice}
Zimmerman,~M.~I.; Porter,~J.~R.; Sun,~X.; Silva,~R.~R.; Bowman,~G.~R.
  \emph{Journal of chemical theory and computation} \textbf{2018}, \emph{14},
  5459--5475\relax
\mciteBstWouldAddEndPuncttrue
\mciteSetBstMidEndSepPunct{\mcitedefaultmidpunct}
{\mcitedefaultendpunct}{\mcitedefaultseppunct}\relax
\EndOfBibitem
\bibitem[Shamsi \latin{et~al.}(2018)Shamsi, Cheng, and
  Shukla]{shamsi2018reinforcement}
Shamsi,~Z.; Cheng,~K.~J.; Shukla,~D. \emph{The Journal of Physical Chemistry B}
  \textbf{2018}, \emph{122}, 8386--8395\relax
\mciteBstWouldAddEndPuncttrue
\mciteSetBstMidEndSepPunct{\mcitedefaultmidpunct}
{\mcitedefaultendpunct}{\mcitedefaultseppunct}\relax
\EndOfBibitem
\bibitem[Lai and Robbins(1985)Lai, and Robbins]{lai1985asymptotically}
Lai,~T.~L.; Robbins,~H. \emph{Advances in applied mathematics} \textbf{1985},
  \emph{6}, 4--22\relax
\mciteBstWouldAddEndPuncttrue
\mciteSetBstMidEndSepPunct{\mcitedefaultmidpunct}
{\mcitedefaultendpunct}{\mcitedefaultseppunct}\relax
\EndOfBibitem
\bibitem[Auer(2002)]{auer2002using}
Auer,~P. \emph{Journal of Machine Learning Research} \textbf{2002}, \emph{3},
  397--422\relax
\mciteBstWouldAddEndPuncttrue
\mciteSetBstMidEndSepPunct{\mcitedefaultmidpunct}
{\mcitedefaultendpunct}{\mcitedefaultseppunct}\relax
\EndOfBibitem
\bibitem[Doerr \latin{et~al.}(2016)Doerr, Harvey, No{\'e}, and
  De~Fabritiis]{doerr2016htmd}
Doerr,~S.; Harvey,~M.; No{\'e},~F.; De~Fabritiis,~G. \emph{Journal of chemical
  theory and computation} \textbf{2016}, \emph{12}, 1845--1852\relax
\mciteBstWouldAddEndPuncttrue
\mciteSetBstMidEndSepPunct{\mcitedefaultmidpunct}
{\mcitedefaultendpunct}{\mcitedefaultseppunct}\relax
\EndOfBibitem
\bibitem[Loncharich \latin{et~al.}(1992)Loncharich, Brooks, and
  Pastor]{loncharich1992langevin}
Loncharich,~R.~J.; Brooks,~B.~R.; Pastor,~R.~W. \emph{Biopolymers: Original
  Research on Biomolecules} \textbf{1992}, \emph{32}, 523--535\relax
\mciteBstWouldAddEndPuncttrue
\mciteSetBstMidEndSepPunct{\mcitedefaultmidpunct}
{\mcitedefaultendpunct}{\mcitedefaultseppunct}\relax
\EndOfBibitem
\bibitem[P{\'e}rez-Hern{\'a}ndez \latin{et~al.}(2013)P{\'e}rez-Hern{\'a}ndez,
  Paul, Giorgino, De~Fabritiis, and No{\'e}]{perez2013identification}
P{\'e}rez-Hern{\'a}ndez,~G.; Paul,~F.; Giorgino,~T.; De~Fabritiis,~G.;
  No{\'e},~F. \emph{The Journal of chemical physics} \textbf{2013}, \emph{139},
  07B604\_1\relax
\mciteBstWouldAddEndPuncttrue
\mciteSetBstMidEndSepPunct{\mcitedefaultmidpunct}
{\mcitedefaultendpunct}{\mcitedefaultseppunct}\relax
\EndOfBibitem
\bibitem[Sutton and Barto(2018)Sutton, and Barto]{sutton2018reinforcement}
Sutton,~R.~S.; Barto,~A.~G. \emph{Reinforcement learning: An introduction};
  2018\relax
\mciteBstWouldAddEndPuncttrue
\mciteSetBstMidEndSepPunct{\mcitedefaultmidpunct}
{\mcitedefaultendpunct}{\mcitedefaultseppunct}\relax
\EndOfBibitem
\bibitem[Pan and Roux(2008)Pan, and Roux]{pan2008building}
Pan,~A.~C.; Roux,~B. \emph{The Journal of chemical physics} \textbf{2008},
  \emph{129}, 064107\relax
\mciteBstWouldAddEndPuncttrue
\mciteSetBstMidEndSepPunct{\mcitedefaultmidpunct}
{\mcitedefaultendpunct}{\mcitedefaultseppunct}\relax
\EndOfBibitem
\bibitem[Piana \latin{et~al.}(2011)Piana, Lindorff-Larsen, and
  Shaw]{piana2011robust}
Piana,~S.; Lindorff-Larsen,~K.; Shaw,~D.~E. \emph{Biophysical journal}
  \textbf{2011}, \emph{100}, L47--L49\relax
\mciteBstWouldAddEndPuncttrue
\mciteSetBstMidEndSepPunct{\mcitedefaultmidpunct}
{\mcitedefaultendpunct}{\mcitedefaultseppunct}\relax
\EndOfBibitem
\bibitem[Waskom()]{kdeplot}
Waskom,~M. {seaborn.kdeplot}.
  \url{https://seaborn.pydata.org/generated/seaborn.kdeplot.html/}, [Online;
  accessed 11-January-2020]\relax
\mciteBstWouldAddEndPuncttrue
\mciteSetBstMidEndSepPunct{\mcitedefaultmidpunct}
{\mcitedefaultendpunct}{\mcitedefaultseppunct}\relax
\EndOfBibitem
\bibitem[Kubelka \latin{et~al.}(2006)Kubelka, Chiu, Davies, Eaton, and
  Hofrichter]{kubelka2006sub}
Kubelka,~J.; Chiu,~T.~K.; Davies,~D.~R.; Eaton,~W.~A.; Hofrichter,~J.
  \emph{Journal of molecular biology} \textbf{2006}, \emph{359}, 546--553\relax
\mciteBstWouldAddEndPuncttrue
\mciteSetBstMidEndSepPunct{\mcitedefaultmidpunct}
{\mcitedefaultendpunct}{\mcitedefaultseppunct}\relax
\EndOfBibitem
\bibitem[Wan and Voelz(2020)Wan, and Voelz]{wan2020adaptive}
Wan,~H.; Voelz,~V.~A. \emph{The Journal of Chemical Physics} \textbf{2020},
  \emph{152}, 024103\relax
\mciteBstWouldAddEndPuncttrue
\mciteSetBstMidEndSepPunct{\mcitedefaultmidpunct}
{\mcitedefaultendpunct}{\mcitedefaultseppunct}\relax
\EndOfBibitem
\end{mcitethebibliography}

\end{document}